\documentclass[pss]{wiley2sp} 
\usepackage{amsmath}
\usepackage{bm}              
\usepackage{w-greek}         

\begin{document}

\title{Engineering molecular aggregate spectra}

\titlerunning{Aggregate spectra}

\author{%
  V. A. Malyshev\textsuperscript{\textsf{\bfseries 1}},
  A. V. Malyshev\textsuperscript{\textsf{\bfseries 1,2,\Ast}}
}
\authorrunning{V. A. Malyshev et al.}

\mail{e-mail
  \textsf{a.malyshev@rug.nl}, Phone
  +31-50-3634784, Fax +31-50-3634754}

\institute{%
  \textsuperscript{1}\,Centre for Theoretical Physics and Zernike
  Institute for Advanced Materials, Nijenborgh 4, 9747 AG Groningen,
  The Netherlands\\
  \textsuperscript{2}\,Ioffe Physical Technical Institute, 194021
  Saint-Petersburg, Russia}

\received{XXXX, revised XXXX, accepted XXXX} 
\published{XXXX} 

\pacs{  42.70.-a    
        67.30.ht    
        68.47.Pe    
        73.20.Mf    
        74.25.Jb    
        8107.-b     
} 

\abstract{%
%
%
%
We show that optical properties of linear molecular aggregates
undergo drastic changes when aggregates are deposited on a metal
surface.
The dipole-dipole interactions of monomers with their images can result in
strong
re-structuring of both the exciton band and the absorption
spectrum, depending on the arrangement of the monomer transition dipoles with
respect to the surface.}

%
%

\maketitle   

\section{Introduction}

Molecular aggregates, in spite of more than 70 years of history after
their discovery by Jelley~\cite{Jelley36} and
Scheibe~\cite{Scheibe37}, still attract a great deal of attention
due to their extraordinary optical and transport properties: narrow
absorption band, superradiant emission, giant higher order
susceptibilities and light harvesting~\cite{Kobayashi96,Knoester02}.
Nowadays, the problem of controlling the optical properties of
molecular aggregates is a challenging task.

In an isolated aggregate, the energy of optical transition is determined by
the transition energy of a monomer, which is fixed for a particular monomer,
as well as by the transfer (dipolar) interactions, which are also fixed by
the aggregate morphology. Therefore, there is almost no way to influence the
optical absorption for a single aggregate. Being embedded into a host, an
aggregate interact with it, which results in the so-called ``solvent'' shift
of its absorption line.  Difference in local solvent environment gives rise
to an inhomogeneously broadened absorption band. Coupling of aggregates to
vibrational degrees of freedom leads to homogeneous broadening of the
aggregate absorption band. In different hosts, both the position of the
aggregate absorption peak and its width vary to some extent. However, both
effects are usually not very pronounced (see, e.g., Ref.~\cite{Renge97}).


As it has been pointed out in a number of papers, the optical
response of J-aggregates changes substantially when J-aggregates are 
deposited on colloidal noble metal nanoparticles. Thus, super-radiant 
lasing at low threshold~\cite{Akins97} and re-structuring of the 
absorption spectra~\cite{Kometani01,Wurtz04,Uwada07,Kelley07}) have 
been reported. We consider aggregates deposited on a flat metallic
surface and show that the aggregate absorption spectrum experience 
strong modifications: the bare absorption band shifts and new bands 
also arise. The effect originates from the interaction of monomer 
transition dipoles with their images in the metal. As a result, a 
simple linear aggregate converts into an effective double-strand 
structure. The changes in the absorption spectrum are sensitive to 
the arrangement of the transition dipoles with respect to the metal 
surface.

Throughout the paper, we restrict ourselves to the case of molecular
J-aggregates. An analogous study for the other type of aggregates,
the so-called H-aggregates deposited on a metal surface, which is 
also an interesting example, will be published elsewhere.


\section{Modeling aggregates}

\subsection{A stand-alone aggregate}

We model a single aggregate as an open linear chain of $N$ optically
active two-level units (monomers) with parallel transition dipoles
coupled to each other by dipole-dipole transfer
interactions which delocalize the excitation. Because of that coupling,
optical excitations of the chain are Frenkel excitons described by the
Hamiltonian
\begin{equation}
    H_{\mathrm{ex}} = \sum_{n=1}^N \> \epsilon_n |n\rangle \langle n| +
    \sum_{n,m}^N\> J_{nm} \> |n\rangle \langle m| \ ,
\label{Hex}
\end{equation}
where $|n \rangle$ denotes the state in which only $n$th monomer is
excited and $\epsilon_n$ is its excitation energy. These energies
are uncorrelated and taken at random from a box distribution
with the mean $\epsilon_0$ (the excitation energy of an isolated
monomer which we set to zero from now on) and the standard deviation
$\sigma$. The dipolar transfer interactions $J_{nm} = J/|n-m|^{3}$
\, $(J_{nn} \equiv 0)$ are considered to be non fluctuating. The
parameter $J = ({\bf d}^2/a^3) (1 - 3\cos^2{\theta})$ represents the
nearest-neighbor transfer interaction, ${\bf d}$ being the
transition dipole moment of a monomer tilted at an angle $\theta$
with respect to the aggregate axis and $a$ being the nearest-neighbor
distance. A negative sign of the couplings $J_{nm}$ corresponds to 
the case of J-aggregates. The exciton energies $\varepsilon_\nu$ \, 
($\nu = 1,\ldots , N$) and wavefunctions $\varphi_{\nu n}$ (taken to 
be real) are obtained after diagonalization of the $N \times N$ 
Hamiltonian matrix $\langle n| H_{\mathrm{ex}} |m \rangle$.

\subsubsection{A single aggregate deposited on a metallic surface}

\begin{figure}[t]%
\begin{center}
\includegraphics*[clip,width=.3\textwidth]{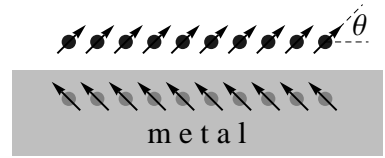}
\end{center}
\caption{
Cartoon of an aggregate deposited on a metal surface. The monomer transition
dipoles (black circles with arrows) induce their images in the metal (gray
circles with arrows). An image is formed in such a way that the component of
the transition dipole perpendicular to the metal surface is conserved, while
the one along the surface is inverted. Real transition dipoles interact with
all images while images do not interact with each other (see the discussion
in the text).
}
\label{AggregateOnSurface}
\end{figure}

We address an aggregate deposited on a conductor surface within the
framework of the image potential approximation, the simplest way to account
for the presence of the metal (see Fig.~\ref{AggregateOnSurface}). The
applicability of the method in the present case can be rationalized as
follows. First, the typical frequency of plasmon oscillations in a good
conductor is on the order of $10^{16} \text{s}^{-1}$, whereas the optical
transition frequency in J-aggregates is about an order of magnitude smaller.
Furthermore, the electron relaxation in metals is much faster than the
exciton optical dynamics. Typical electron relaxation time is $10\div100$ fs
(see, e.g., Ref.~\cite{Palik85}), while the characteristic optical dynamics
time in the prototype J-aggregates of the pseudoisocyanine dye is
$10\div100$ ps (see, e.g., Refs.~\cite{Fidder90,Heijs05}). Therefore,
conduction electrons represent a fast system instantly following all charge
redistributions in the aggregate at the exciton related time scale.

It is worthwhile to note also that, contrary to the present case of a
\emph{bulk} metallic substrate, the image potential method can usually not
be justified for a J-aggregate adsorbed onto a metal \emph{nano-particle}.
In the latter case, the frequency of surface plasmons in the particle is
often on the order of the optical frequency, therefore strong mixing of
excitons and plasmons should be taken into account (see discussions in
Refs.~\cite{Larkin04} and~\cite{Markel05}).

Within the framework of the image approximation, the total Hamiltonian
$H = H_{\mathrm{ex}} + H_{\mathrm{img}} + H_{\mathrm{ex-img}}$ consists of
the Hamiltonian of the aggregate $H_{\mathrm{ex}}$, Eq.~(\ref{Hex}), that of
the image dipoles $H_{\mathrm{img}}$ and the Hamiltonian
$H_{\mathrm{ex-img}}$ which describes the interaction of the aggregate
monomers with all images in the metal. It is straightforward to write out
explicit expressions for $H_{\mathrm{img}}$ and $H_{\mathrm{ex-img}}$. We
note that (i) the image Hamiltonian in the cite representation is diagonal
with image transition energies equal to those of the corresponding monomers,
and that (ii) image dipoles do not interact with each other. The latter can
be understood in the following way. A charge near a metal surface gives rise
to the redistribution of charges on the surface, which is equivalent to
introducing an image charge. The surface charge distribution created by two
charges is a sheer linear combination of the two distributions induced by
the charges independently (because of the superposition principle). Should
there be an interaction between the corresponding images, the overall
potential would have an additional part due to the interaction, which is not
the case. However, the interaction of images with the external
electromagnetic field should be taken into account. The following comment is
due here: consider radiation field of a dipole emitter near a metal surface;
the field of the system in the far zone would contain the contribution from
the dipole and the one from the induced surface charge distribution, the
latter being equivalent to the field of the image. The system would emit
therefore as a quadrupole, suggesting that we should treat it as a system of
two real dipoles in this case. The same applies to the excitation of the
system by an external field.

The new exciton energies and wavefunctions can be obtained after
diagonalization of the $2N \times 2N$ Hamilton matrix $\langle n| H |m
\rangle$, similar to the case of the stand-alone aggregate.

\begin{figure}[t]%
\begin{center}
\includegraphics*[clip,width=0.4\textwidth]{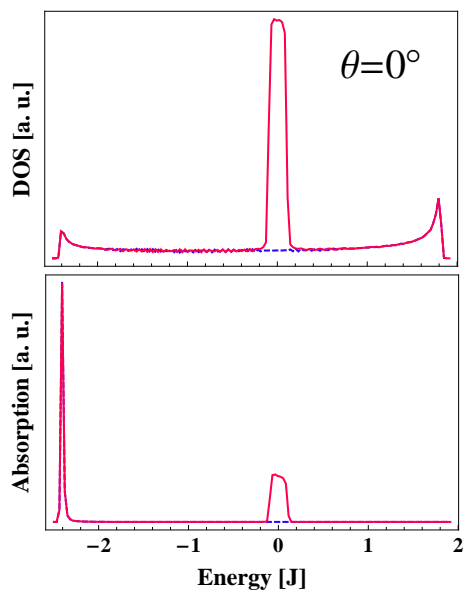}
\end{center}
\caption{
DOS (upper panel) and absorption spectra (lower panel) of aggregates in a
glassy host (dashed curves) and deposited on a metallic surface (solid
curves). The set of parameters used in simulations is: the disorder strength
$\sigma = 0.2 J$, the angle between the monomer transition dipole and the
surface $\theta = 0$, and the distance between aggregates and the surface $d
= a$.
}
\label{DOS_Iabs_on_surface_d=2_theta=0_fi=0}
\end{figure}

\section{Results and discussion}

We calculated the density of states (DOS) and the
absorption spectrum defined, respectively, as
\begin{equation}
    \mathrm{DOS}(E)  = \frac{1}{2N} \Bigg\langle \sum_{\nu=1}^{2N} 
    \delta(E - \varepsilon_{\nu}) \Bigg\rangle
\label{DOS}
\end{equation}
\begin{equation}
A(E) = \frac{1}{2N} \Bigg\langle \sum_{\nu=1}^{2N} \Bigg (\sum_{n=1}^{2N}
\varphi_{\nu n} \Bigg )^2 \delta(E - \varepsilon_{\nu}) \Bigg\rangle \ ,
\label{A}
\end{equation}
where the angular brackets denote the average over disorder realizations. In
the simulations, aggregates of $N = 300$ monomers were considered. For
aggregates deposited on a metal surface, we assumed the transition dipole
moments of monomers $\bf{d}$ lying in the plane perpendicular to the metal
surface.
%

\begin{figure}[t]%
\begin{center}
\includegraphics*[width=0.4\textwidth]{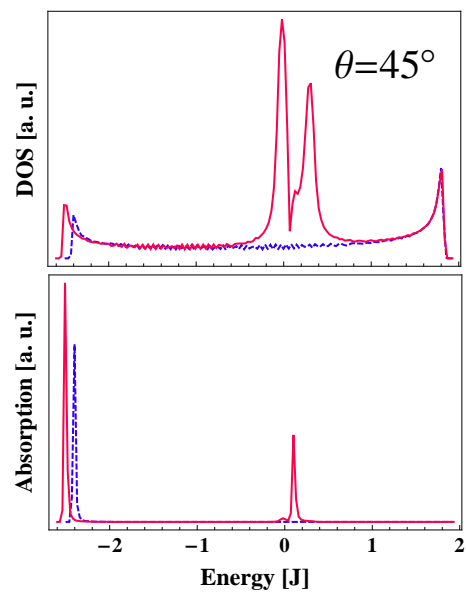}
\end{center}
\caption{
Same as in Fig.~\ref{DOS_Iabs_on_surface_d=2_theta=0_fi=0} calculated for
$\theta = \pi/4$.
}
\label{DOS_Iabs_on_surface_d=2_theta=45_fi=0}
\end{figure}

In Figs.~\ref{DOS_Iabs_on_surface_d=2_theta=0_fi=0}
and~\ref{DOS_Iabs_on_surface_d=2_theta=45_fi=0} we show the results of our
simulations of the DOS (upper panel) and the absorption spectra (lower
panel) for $\sigma=0.2J$. The dashed curves represent the data obtained for
stand-alone aggregates in a glassy host (in the absence of the metal). The
solid curves show the same quantities calculated for aggregates deposited on
the surface of a metal with the aggregate-surface separation $d = a$, so
that the distance between a monomer and its image is equal to $2a$. The data
presented in Figs.~\ref{DOS_Iabs_on_surface_d=2_theta=0_fi=0}
and~\ref{DOS_Iabs_on_surface_d=2_theta=45_fi=0} were obtained for the
tilting angles $\theta = 0$ and $\theta = \pi/4$, respectively.

In the absence of the metal, the DOS and the absorption spectra exhibit
standard shapes: the DOS reveals two singularities at the band edges
(smoothed by the disorder) while the absorption has a peak at the lower band
edge (the J-band). The picture changes dramatically for deposited
J-aggregates. Along with the band edge singularities of the DOS, very
pronounced features arise at the center of the band for both tilting angles
(upper panels in Figs.~\ref{DOS_Iabs_on_surface_d=2_theta=0_fi=0}
and~\ref{DOS_Iabs_on_surface_d=2_theta=45_fi=0}). The DOS reconstruction is
accompanied by the occurence of an additional intense line at the band
center (about the monomer energy). For the tilting angle $\theta = \pi/4$
the bare J-band shifts to the red, reflecting the corresponding shift of the
lower energy band edge.

The changes that occur in the DOS and the absorption spectra of deposited
aggregates can be qualitatively understood within a simplified model which
takes into account the coupling of a monomer to its own image only,
neglecting couplings to images of other monomers. A deposited aggregate can
then be viewed as a system of dimers (monomer-image) coupled to each other
via the monomer-monomer transfer interactions $J_{nm}$. For $d=a$ and
$\theta=0$, the magnitude of coupling between neighboring monomers, $|J| =
2|{\bf d}|^2/a^3$, is 16 times larger than that between a monomer and its
image, $|{\bf d}|^2/8a^3$. In this case the images can be considered to be
almost decoupled from the monomers, which gives rise to $\delta$-peaks in
the DOS at monomer energies $\varepsilon_n$. The image states would have the
monomer oscillator strength, resulting in the appearance of the new
absorption band about the monomer energy (see
Fig.~\ref{DOS_Iabs_on_surface_d=2_theta=0_fi=0}). The lineshape of the new
band reproduces the distribution function of the on-site energies. In the
vicinity of the ``magic'' angle $\theta=\arccos(1/\sqrt{3}) \approx
57^{\circ}$, at which the monomer-monomer transfer interactions $J_{nm}$
vanish, the system can be viewed as a set of weakly coupled dimers. The two
eigenstates of a dimer would then give rise to their own bands due to weak
coupling to neighbors. This trend can be seen in the
Fig.~\ref{DOS_Iabs_on_surface_d=2_theta=45_fi=0} where the appearance of
band edge singularities in the DOS is clearly visible. The singularities
give rise to additional features in the absorption. A more comprehensive
study of the optical properties of deposited molecular aggregates will be
published elsewhere.


\section{Summary and concluding remarks}

We studied numerically the absorption spectra of J-aggregates deposited on a
surface of a metal. We found that the density of states and the absorption
spectra undergo strong changes which depend on the arrangement of the
monomer transition dipoles with respect to the metal surface. In particular,
within some range of the transition dipole tilting angle, the original
J-band shifts to the red and a new intense absorption band appears in the
middle of the exciton band (about the monomer energy). Our findings open a
new perspective for designing optical properties of molecular aggregates.

\begin{acknowledgement}
A. V. M. acknowledges support from NanoNed, a national nanotechnology
program coordinated by the Dutch Ministry of Economic Affairs.
\end{acknowledgement}

%
%

\end{document}